\documentclass[12pt]{iopart}
\usepackage{graphicx}
\usepackage[normal]{caption2}
\begin{document}
\jl{3}

\title[]
{Has the FFLO state been observed in the organic 
superconductor $\kappa-$(BEDT-TTF$)_2$Cu(NCS$)_2$  ?}

\author{S Manalo and U Klein\footnote[3]{To
whom correspondence should be addressed.}}

\address{Johannes Kepler Universit{\"a}t Linz, Institut f{\"u}r Theoretische 
Physik, A-4040 Linz, Austria}

\begin{abstract}
We compare the theoretical anisotropic upper critical field $H_{C}(\Theta,T)$ of 
a quasi-two-dimensional d-wave superconductor with recent $H_{c2}$ data for the 
layered organic superconductor $\kappa-(BEDT-TTF)_2Cu(NCS)_2$. We find agreement 
both with regard to the angular and the temperature dependence of $H_{C}$. 
This supports the suggestion that the Fulde-Ferrell-Larkin-Ovchinnikov 
state (FFLO state) exists in this material for exactly plane-parallel 
orientation of the magnetic field. Indications of precursor states, 
occurring for small deviations from the plane-parallel field direction, 
are also pointed out and further measurements for confirming the 
existence of the FFLO state are proposed. 
\end{abstract}


If the superconducting state in high magnetic fields is limited by paramagnetic
pair breaking alone, the transition from the homogeneous
superconducting state to the normal 
conducting state may proceed  either directly, at the Pauli limiting field 
$H_P$, or via an interposed inhomogeneous superconducting state predicted 
in 1964 by Fulde and Ferrell~[1] and by Larkin and Ovchinnikov~[2] (FFLO state). 
The latter state, which stabilizes as a consequence of spin polarization, has 
attracted considerable interest over the years, but has, despite of this, not 
yet been definitely verified experimentally.

Favourable conditions for observing the FFLO state are found in clean
superconductors with orbital critical fields much larger than $H_P$. 
In practice, it seems always necessary to reduce the orbital pair breaking 
effect by using layered superconductors with nearly decoupled planes or 
extremely thin films (quasi-two-dimensional superconductors) and applying the 
magnetic field in a direction parallel to the conducting planes. 
Several classes of superconducting materials with favourable conditions for 
observing the FFLO state do exist. These include the ``classical'' intercalated 
transition metal dichalcogenides as well as more exotic  
materials like High-$T_c$ compounds and organic superconductors.

In this Letter we refer to a recent measurement of the upper critical field 
$H_{c2}$ in the organic superconductor $\kappa-(BEDT-TTF)_2Cu(NCS)_2$~[3]. This  
layered material shows strong anisotropy of the superconducting properties with 
regard to out-of-plane directions. In addition, a number of experiments
listed in reference [3] are interpreted in terms of a strong 
in-plane anisotropy, of d-wave type, of the gap parameter. The coherence 
length $\xi_\perp$ perpendicular to the layers of this clean material is smaller 
than the interlayer spacing $d$ and one expects an extreme reduction of 
orbital pair breaking for plane-parallel applied field. 
The angular dependence of $H_{c2}$ was measured in Ref~[3] both with respect to 
the angle $\Theta$ between applied magnetic field and the direction normal to 
the conducting planes and with respect to the azimuthal angle $\phi$, which denotes 
the direction of magnetic field lying within the plane. The results showed, as 
expected, a strong variation of $H_{c2}$ with $\Theta$. On the other hand, no dependence 
on $\phi$ was observed. The maximal value of $H_{c2}$ at the plane-parallel position 
$\Theta=\pi/2$ was of the order of, but ~50\% higher than, the Pauli paramagnetic 
limit $H_P$. These facts led the authors of reference [3] to propose that their 
in-plane critical field is the phase boundary between the normal-conducting 
state and the FFLO state of a d-wave superconductor.

We examined this question by calculating the angular 
and temperature dependence of the theoretical phase boundary $H_C(\Theta,T)$
between the normal-conducting state and the superconducting states  of a 
quasi-two-dimensional d-wave superconductor. Comparison with the data of 
reference [3] showed good agreement, supporting the hypothesis  that in this 
experiment the phase boundary of a FFLO state (for d-wave superconductors) 
has been observed for the first time.

We assume that the coupling between the conducting planes of 
$\kappa-(BEDT-TTF)_2Cu(NCS)_2$ can be neglected, making our problem effectively 
two-dimensional. Then, if the field has both a perpendicular and a parallel 
component, the superconducting state is limited by both orbital and 
paramagnetic pair breaking. For exactly plane-parallel magnetic field 
the FFLO phase should be realized below a reduced temperature ($t=T/T_c$) 
of $t\approx0.4$~[4]. Such a situation, with competition between both pair breaking 
effects, has been studied first  by Bulaevskii~[5]. His treatment was 
later generalized to arbitrary temperatures and d-wave superconductors by 
Shimahara and Rainer~[6]. Below the stability limit of the normal-conducting
state, which will be referred to as $H_C(\Theta,T)$, a series of different 
inhomogeneous superconducting states, depending on $\Theta$, appear. For s-wave 
superconductors each one of these states belongs to a particular value of 
Landau's quantum number $n$, which takes integer values $n=0,1,2,..$.
These states are (for s-wave) the following:
\begin{itemize}
\item The vortex state for small $\Theta$ belongs to $n=0$.
\item A series of inhomogeneous states for $\Theta$ near $\pi/2$, each one 
characterized by a single value $n>0$, with $n$ increasing with increasing $\Theta$.
\item The FFLO state for $\Theta=\pi/2$, which may be characterized by $n \to \infty$. 
\end{itemize}
The structure of the higher Landau level states, for $n>0$, has been calculated 
recently for s-wave superconductors by minimizing the quasiclassical free 
energy~[7]. For d-wave superconductors~[6] a state below $H_C(\Theta,T)$ is no longer 
characterized by a single value of $n$ but rather by an infinite subset 
$\{n_0,n_0 \pm 4,n_0 \pm 8,\ldots\}$. However, the dominant contribution may still be 
characterized by a single number $n$, which increases again with increasing  
$\Theta$ and approaches infinity in the FFLO limit. Thus, basically the above 
classification scheme remains valid for d-wave symmetry. The phase boundary 
of the `pure' FFLO state for d-wave superconductors [the curve $H_C(\pi /2,T)$] has 
been calculated by Maki and Won~[4].

The linearized gap equation to be solved is given by~[5,6]
\begin{eqnarray}
     -\log(\frac{T}{T_c}) \Delta(\vec{r}) )
       = \pi k_B T \int_0^{\infty} \frac{ds}{\sinh(\pi k_B T s)} 
       \int_0^{2\pi}\frac{d \phi'}{2 \pi} 
       \left[ \gamma(\hat{p}^\prime)^2] \right][1 \nonumber\\
       - \cos[s\{\mu_0H - \frac{1}{2} \vec{v}_F^\prime \vec{\Pi}\}] ]\Delta(\vec{r})
\mbox{.} 
\end{eqnarray}
We consider a cylindrical Fermi surface, appropriate for the present 
two-dimensional problem, with the Fermi velocity $\vec{v}_F=v_F(\vec{e}_x\cos\phi
+\vec{e}_y\sin{\phi})$. The gap parameter is given by 
$\Delta(\vec{r},\hat{p})=\Delta(\vec{r}) \gamma(\hat{p})$, where $\gamma(\hat{p})=1$ for s-wave, 
and $\gamma(\hat{p})=\sqrt{2} (\hat{p}_ x^2-\hat{p}_ y^2)=1+\cos(4 \phi)$ for d-wave 
pairing. The canonical momentum is defined by  $\vec{\Pi}=
\frac{\hbar}{\imath}(\nabla-\imath \frac{2e}{\hbar c}\vec{A})$. The magnetic field 
$\vec{H}$ is assumed to lie in the $yz$-plane, with $H_y=H_\parallel=H\sin\Theta$ 
and $H_z=H_\perp=H\cos\Theta$. We use the following gauge for the vector 
potential:  $A_x=H_\parallel z-H_\perp y \mbox{, and } A_y=A_z=0$. Paramagnetic pair breaking
enters via the term $\mu_0H$ in (1); the electron's magnetic moment is $\mu_0=-g_L \mu_B /2$, 
with the Lande factor $g_L\approx 2$ and Bohr`s magneton $\mu_B=\hbar e/(2mc)$.
The method of reducing equation (1) to a set of algebraic equations follows 
exactly reference [6] and need not be repeated here. As a first check of our 
numerical method we compared our results with reference [6] and found complete 
agreement whenever the same set of input parameters was used. 

Let us proceed to a comparison of the solutions $H_{C}$ of equation (1) with the 
$H_{c2}$ data reported in reference [3]. The first point to address
is the independence of $H_{c2}$ on the azimuthal angle $\phi$. Such a  
dependence may easily be incorporated in the present model by replacing 
$\vec{v}_F^\prime$ in (1) by $v_F\left[\vec{e}_x\cos(\phi^\prime + \phi) +\vec{e}_y\sin(\phi^\prime + \phi)\right]$. 
As long as orbital pair-breaking is present for a parallel field component, 
the symmetry-breaking term $A_x=H_\parallel z$ has to be kept in (1). If, on the other 
hand, complete decoupling of the (infinitely thin) conducting planes can be 
assumed, the term   $H_\parallel z$ can be dropped and $H_{C}$ becomes independent of $\phi$, 
as can be explicitely confirmed numerically. In this context, we recall that 
a three-dimensional d-wave superconductor shows anisotropy of the  upper 
critical field~[8]. Thus, the observed independence of $H_{c2}$ on $\phi$ is in 
agreement with the present model and, in particular, with the assumption of 
quasi-two-dimensional superconductivity.

If $T$ is measured in units of $T_c$ and $H$ in units of $\mu \Delta_0$ (where $\Delta_0$ is the 
BCS-gap at $T=0$), the present model requires only one single parameter 
$k_BT_c/E_F$ to be fitted. This parameter is proportional to the ``bulk'' ratio 
of the orbital and spin critical fields $\hbar c /(2 e \xi_0^2)$ and $k_BT_c  / \mu_0$ 
respectively. In the present anisotropic model the actual ratio of spin and 
orbital pair-breaking depends on $\Theta$ and may be written in the form 
$k_BT_c/(E_F\cos\Theta)$. 
The best fit to the  $\Theta$-dependence of $H_{c2}$ at $T=1.45\,K$ 
(see figure 4 of reference [3]) has been obtained for $k_BT_c/E_F=0.058$. This 
value of $k_BT_c/E_F$ is consistent with a critical 
temperature $T_c=10.4\,K$ of the sample studied in reference [3], and 
a Fermi energy 
of the order of $100\,K$ as estimated from several experiments~[9]. Using this 
value we found very good agreement between $H_C$ and the data of 
reference [3], as shown in figure 1. The theoretical curves in figure 1 have 
been calculated assuming d-wave symmetry; the difference between d-wave and 
s-wave was found to be small except very close to $T=0\,K$ and 
$\Theta=90 \mbox{ deg}$.
\begin{figure}[htbp]
  \begin{center}
\includegraphics[angle=270,width=13cm]{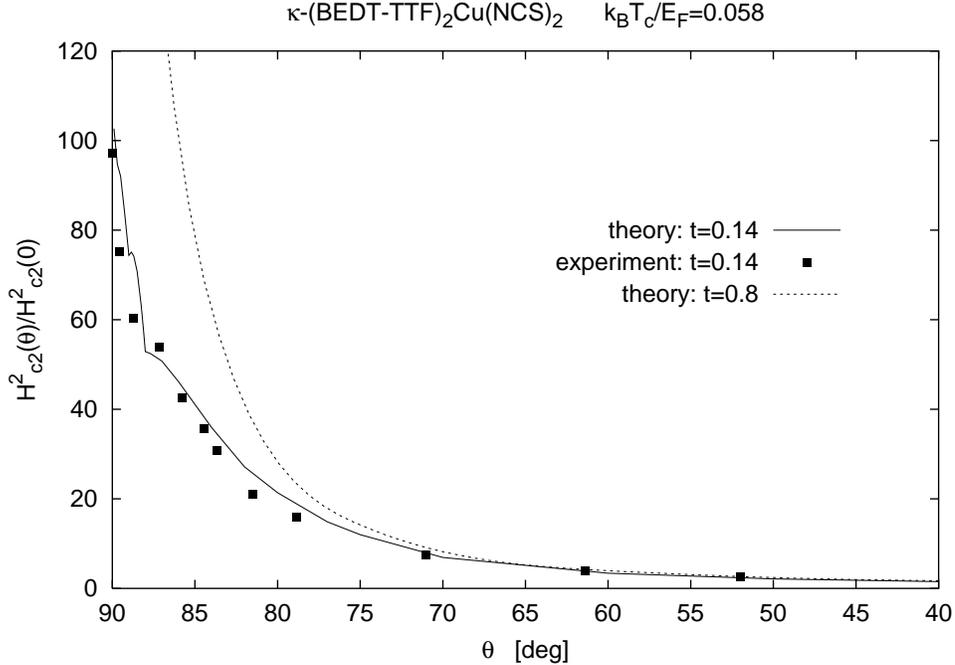}    
    \caption{Square of upper critical field, normalized to its value at 
$\Theta=0$, as a function of $\Theta$. Full squares: data 
of reference [3] at $t=T/T_c=0.14$. Full line: theoretical result for the 
Bulaevskii-Shimahara-Rainer phase boundary for $t=0.14$. 
Dashed line: theoretical result for $t=0.8$.}
    \label{fig:Fig.1}
  \end{center}
\end{figure}

The $H_{c2}$ data show a small but clearly visible kink near the 
plane-parallel orientation, at $\Theta \approx 87 \mbox{ deg}$. A similar feature is 
also found in the theoretical phase boundary $H_C(\Theta)$, as shown in figure 1 
(where the square of the critical fields has been plotted in order to make 
this discontinuous change in slope better visible). This kink  
indicates the transition from the vortex state, with $n=0$, to the first of 
the above mentioned FFLO-precursor states, with $n=1$. Still closer to 
$\Theta \approx 90 \mbox{ deg}$ equation (1) yields additional transitions corresponding 
to $n=2,3$, which are still visible in figure 1 in the theoretical curve but 
not in the data points. The $H_C$-curves  describing the $n=0,1,2$ 
transitions, for the same material but higher $T$, are shown on a larger 
scale in figure 2. The order parameter structure of the precursor 
phases, where pairing takes place in Landau levels $n>0$, has been 
investigated recently for s-wave superconductors~[7]. Two types 
of such precursor states have been found: (i) quasi-one-dimensional states, 
which may be considered as a mixture of rows of vortices and one-dimensional 
FFLO-type oscillations, and (ii) two-dimensional lattices  with several 
zeros of the order parameter with different vorticity~[7]. Such unusual 
states [of type (ii)] have been predicted to occur in the extremely high 
field region where quantization of single electron levels becomes 
important~[10]. The present arrangement  might provide a relatively feasible 
way of observing such vortex structures. It should be mentioned, however, 
that for d-wave superconductors neither the equilibrium structure of the 
FFLO state nor that of the  $n>0$ states has been calculated so far.
\begin{figure}[htbp]
  \begin{center}
\includegraphics[angle=270,width=13cm]{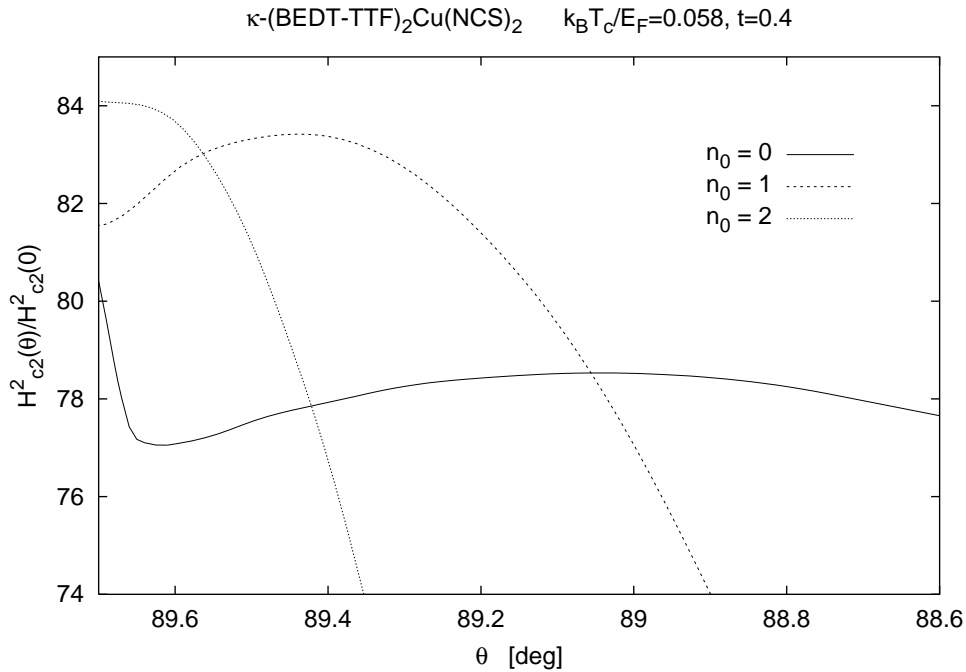}    
    \caption{The branches $n_0=0,1,2$ of the upper critical field, as 
calculated from equation (1), are plotted in more detail, using the same 
value of $k_B T_C/E_F$ as in figure 1 but a higher temperature $t=0.4$.}
    \label{fig:Fig.2}
  \end{center}
\end{figure}
\begin{figure}[htbp]
  \begin{center}
\includegraphics[angle=270,width=13cm]{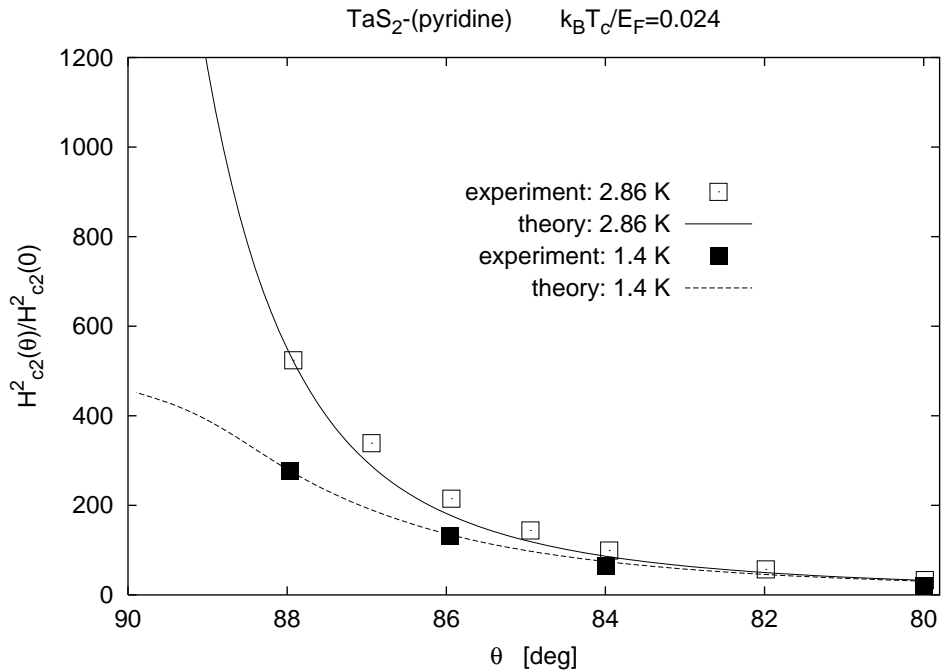}    
    \caption{Comparison of the $\Theta$-dependence of the upper 
critical field of $TaS_2-(pyridine)$, as reported in reference [11], with 
the solutions of equation (1) for s-wave superconductivity at $t=2.86$ and 
$t=1.4$.}
    \label{fig:Fig.3}
  \end{center}
\end{figure}

The shape of the $H_C(\Theta)$ curve depends distinctively on temperature, as 
shown by the plot (dashed line) of $H_C(\Theta)^2$ at $t=0.8$ in figure 1. The 
reason for this enhancement at higher $T$ is, of course, that paramagnetic 
pair breaking becomes less effective at higher temperature. Data at higher 
$T$ have not been reported in reference [3] but would be useful in order to 
check the present interpretation. Looking for similar measurements we found 
old data by Morris and Coleman~[11] for intercalated transition metal 
dichalcogenide $TaS_2-(pyridine)$ samples. In this material, which represents 
a nearly perfect realization of two-dimensional superconductivity, an 
unexplained anomaly with regard to the behavior of the upper critical field 
at different $T$ has been reported (figure 10  of reference [11]). We find 
excellent agreement (see figure 3) comparing the data of reference [11] with the 
solutions of equation (1) for an s-wave superconductor. Again, a single 
parameter has been adjusted ($k_BT_c/E_F=0.024$) to obtain both of the 
theoretical curves shown in figure 3. The resistance data reported in 
reference [11] show a non-monotonic behavior near the plane-parallel field 
orientation (see figure 3 of reference [11]), which may be due to transitions 
to the $n>0$ states. The latter states are discussed in more detail in 
reference [7]. 
\begin{figure}[htbp]
  \begin{center}
\includegraphics[angle=270,width=13cm]{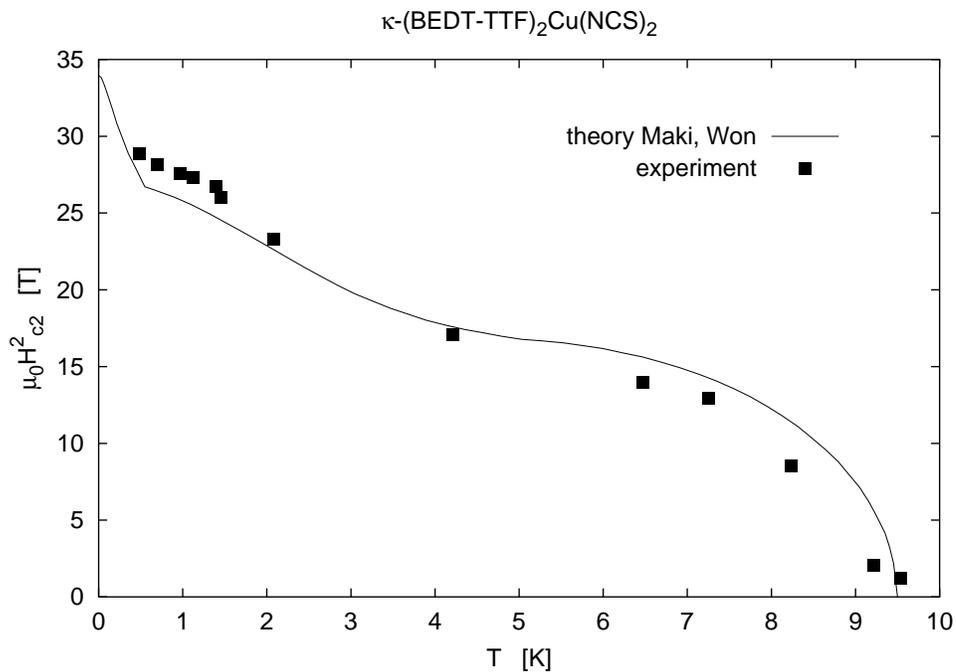}    
    \caption{Comparison of the temperature dependence of the 
plane-parallel upper critical field reported in reference [3] with the 
theoretical result reported in reference [4].}
    \label{fig:Fig.4}
  \end{center}
\end{figure}

Finally, let us compare the measured temperature-dependence of $H_{c2}$ for 
the plane-parallel field orientation (figure 3 of reference [3]) with 
$H_C(\pi /2,T)$. According to our interpretation of these data, the states 
below $H_{c2}(\pi/2,T)$ should be a d-wave version of the FFLO state for 
$T < T^\ast\cong0.4\,T_c$, and the homogeneous superconducting state for $T > T^\ast$. 
This phase boundary has been calculated first by Maki and Won~[4]. The Shimahara 
Rainer d-wave phase boundary must agree with reference [4] in the limit $\Theta\to\pi/2$. 
We found agreement, except for the steep rise of $H_C(\pi /2,T)$ below $0.05 T_c$ 
reported in reference [4] (a possible reason for this discrepancy may be slow 
convergence of our numerical method for low $T$ and high $n$). The comparison 
between theory~[4] and experiment~[3] depicted in figure 4 shows again fairly 
good agreement. A characteristic  difference in temperature variation above and 
below $T^\ast$ is visible in the data points, although it is less pronounced than in 
the theoretical curve. The difference in  
critical fields between s-wave and d-wave is again rather small; at 
$T=0.05\,T_c$, the lowest temperature, where measurements for 
$\kappa-(BEDT-TTF)_2Cu(NCS)_2$ have been reported~[3], both are approximately 
given by the standard $2D$-result~[12] for the FFLO state, $\mu_0H_C=\Delta_0$, which 
exceeds the Pauli limiting field by $\cong40\%$. Thus, these numbers do also fit 
well into the FFLO interpretation of the phase boundary for plane-parallel 
field orientation.

Summarizing, the proposal of Nam et al.~[3] that upper critical field data 
for a plane-parallel field orientation in the layered organic superconductor 
$\kappa-(BEDT-TTF)_2Cu(NCS)_2$ should be interpreted in terms of a FFLO state, has been
supported by our calculations. The data agree with the predictions of a 
model of a quasi-two-dimensional superconductor both with regard to the 
angular and the temperature dependence of the critical field. Further 
confirmation of this interpretation could be obtained by means of measurements 
at higher temperatures, where paramagnetic pair-breaking is strongly reduced. 
If this interpretation is correct, precursor states with interesting properties 
should appear for applied fields close to the plane-parallel orientation.

\section*{References}

\end{document}